\documentclass[preprint,12pt]{elsarticle}

\usepackage{amssymb}
\usepackage{amsmath}
\usepackage{graphicx}%
\usepackage{multirow}%
\usepackage[title]{appendix}%
\usepackage[table]{xcolor}%
\usepackage{textcomp}%
\usepackage{manyfoot}%
\usepackage{booktabs}%
\usepackage{algorithm}%
\usepackage{algorithmicx}%
\usepackage{algpseudocode}%
\usepackage{listings}%
\usepackage{pifont}
\usepackage{amsfonts}
\usepackage{url}
\usepackage{amsthm}%
\usepackage{mathrsfs}%
\usepackage[switch]{lineno}
\usepackage{inconsolata}
\usepackage{bbding}
\usepackage{colortbl}
\usepackage{tabularx}
\usepackage{array}
\usepackage{subcaption}
\usepackage{hyperref}

\journal{Expert System with Applications}

\newcommand{\gt}{\cellcolor{gray!25}}
\newcommand{\rt}{\textcolor[rgb]{0.75,0.25,0.25}}

\newcommand{\lzhmark}[1]{\textcolor{black}{#1}}
\newcommand{\revision}[1]{\textcolor{black}{#1}}

\def \ourdataset{POJ-Evl}
\begin{document}

\begin{frontmatter}

\title{Functional Consistency of LLM Code Embeddings: A Self-Evolving Data Synthesis Framework for Benchmarking}

\author[1]{Zhuohao Li}
\ead{lizhh268@mail2.sysu.edu.cn}
\address[1]{School of Computer Science and Engineering, Sun Yat-Sen University, Guangzhou, 510006, China}

\author[2]{Wenqing Chen \corref{cor1}}
\ead{chenwq95@mail.sysu.edu.cn}
\address[2]{School of Software Engineering, Sun Yat-Sen University, Zhuhai, 519082, China}

\author[3,4]{Jianxing Yu}
\ead{yujx26@mail.sysu.edu.cn}
\address[3]{School of Artificial Intelligence, Sun Yat-Sen University, Zhuhai, 519082, China}
\address[4]{Key Laboratory of Sustainable Tourism Smart Assessment Technology, Ministry of Culture and Tourism of China, Zhuhai, 519080, China}

\author[6]{Zhichao Lu}
\ead{zhichao.lu@cityu.edu.hk}
\address[6]{Department of Computer Science, City University of Hong Kong, Hong Kong, China}
\cortext[cor1]{Corresponding author}

\begin{abstract}
Embedding models have demonstrated strong performance in tasks like clustering, retrieval, and feature extraction while offering computational advantages over generative models and cross-encoders. Benchmarks such as MTEB have shown that text embeddings from large language models (LLMs) capture rich semantic information, but their ability to reflect code\lzhmark{-level functional} semantics remains unclear. Existing studies largely focus on code clone detection, which emphasizes syntactic similarity and overlooks functional understanding. In this paper, we focus on the functional consistency of LLM code embeddings, which determines if two code snippets perform the same function regardless of syntactic differences. We propose a novel data synthesis framework called \lzhmark{Functionality-Oriented Code Self-Evolution} to construct diverse and challenging benchmarks. Specifically, we define code examples across four semantic and syntactic categories and find that existing datasets predominantly capture syntactic properties. Our framework generates four unique variations from a single code instance, providing a broader spectrum of code examples that better reflect functional differences. Extensive experiments on three downstream tasks-code clone detection, code functional consistency identification, and code retrieval-demonstrate that embedding models significantly improve their performance when trained on our evolved datasets. These results highlight the effectiveness and generalization of our data synthesis framework, advancing the functional understanding of code.
\end{abstract}

\end{frontmatter}

\section{Introduction}
Code intelligence tasks, such as code generation~\cite{SIRBU2025125821}, clone understanding~\cite{SHENEAMER2018405,martinez2024source}, and code retrieval~\cite{sachdev2018retrieval,ling2021deep}, are fundamentally concerned with the capability to understand and represent the semantics of code---its functional consistency across different implementations. Accurate evaluation of code segment functionality, regardless of their syntactic differences, is critical for numerous applications in software engineering and beyond \cite{SUDHAMANI201963,CUI2025127470,alon2018codeseq,alon2019code2vec}. However, existing datasets and evaluation benchmarks primarily focus on syntactic similarities, which are insufficient for capturing the nuanced semantic understanding required for these tasks.

Recent advances in large language models (LLMs), such as ChatGPT~\cite{chatgpt}, have demonstrated impressive capabilities to understand and generate human-like code through zero-shot or few-shot prompting. While the progress is impressive, LLMs often struggle to distinguish complex code examples that demand functional understanding beyond superficial syntactic patterns~\cite{li2023nuances}. Moreover, applying generative LLMs or cross-encoders for pairwise code comparison is computationally expensive: analyzing $N$ code samples requires $C^{2}_{N} = \frac{N(N-1)}{2}$ runs due to the pairwise nature of the task. In contrast, embedding models, which map code into numerical vector representations, offer significant computational advantages by processing each code snippet independently in $N$ runs.

\begin{figure*}[t]
    \centering
        \centering
        \includegraphics[width=1.0\textwidth]{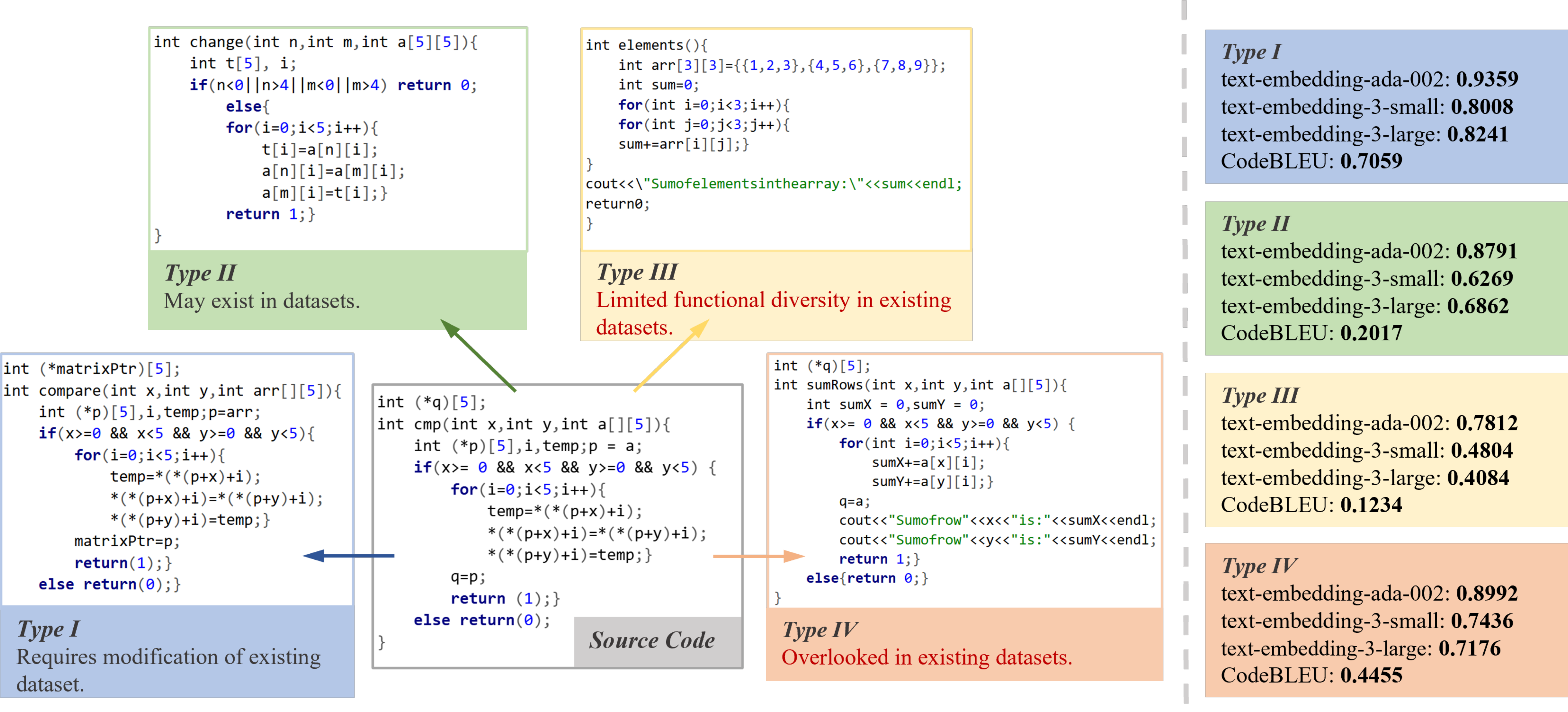}
    \caption{\revision{
    Examples and Distribution of Four Code Types in Current Datasets: Predominantly Type I \& II, Limited Complex Functional Variants (Type III \& IV).
Each variant is labeled with its CodeBLEU score and functional consistency from three embedding models.
While CodeBLEU aligns with type definitions, embedding models still struggle to detect functional consistency, highlighting the challenge posed by our synthesized cases.}}
    \label{fig:agu_arch_a}
    \vspace{-1em}
\end{figure*}

Despite their advantages, LLM embeddings remain underexplored in their ability to capture functional consistency, primarily due to the lack of diverse and comprehensive datasets. Traditional datasets, such as CodeXGLUE~\cite{lu2021codexglue}, often prioritize syntactic similarities for pairwise code comparison. 
\begin{table}[h]
    \caption{
    Four types of code pairing scenarios categorized by syntactic and semantic consistency. The symbols "\Checkmark" and "\small\XSolid" represent consistency and inconsistency, respectively.
    } 
    \centering
    \resizebox{0.6\textwidth}{!}
    {
        \begin{tabular}{ c l |  c  c }
            \toprule
            \multicolumn{2}{c}{\textbf{Code Types}} & \textbf{Syntax} & \textbf{Semantics}  \\ 
            \hline
            \multirow{2}{*}{Positive} & Type I & \Checkmark & \Checkmark \\
            & Type II & \small \XSolid & \Checkmark \\
            \hline
            \multirow{2}{*}{Negative} & Type III & \small \XSolid & \small \XSolid \\ 
            & Type IV & \Checkmark & \small \XSolid \\ 
            \bottomrule            
        \end{tabular}
    }
    \label{tab:types}
\end{table}
As shown in Table~\ref{tab:types}, we categorize code pairs into four types based on syntactic and semantic aspects: Type I and Type II represent positive samples with the same functionality, while Type III and Type IV represent negative samples with differing functionality. Most existing datasets predominantly include Type I and Type II pairs, with limited representation of the more complex functional variations (e.g., Type III and Type IV), as shown in 
Figure~\ref{fig:agu_arch_a}.
For instance, datasets like POJ-104~\cite{poj104} primarily consist of programming tasks with a narrow range of code variations, failing to challenge models to distinguish between functional consistency and syntactic similarity.

\begin{figure}[tb]
  \centering
  \includegraphics[width=.65\textwidth]{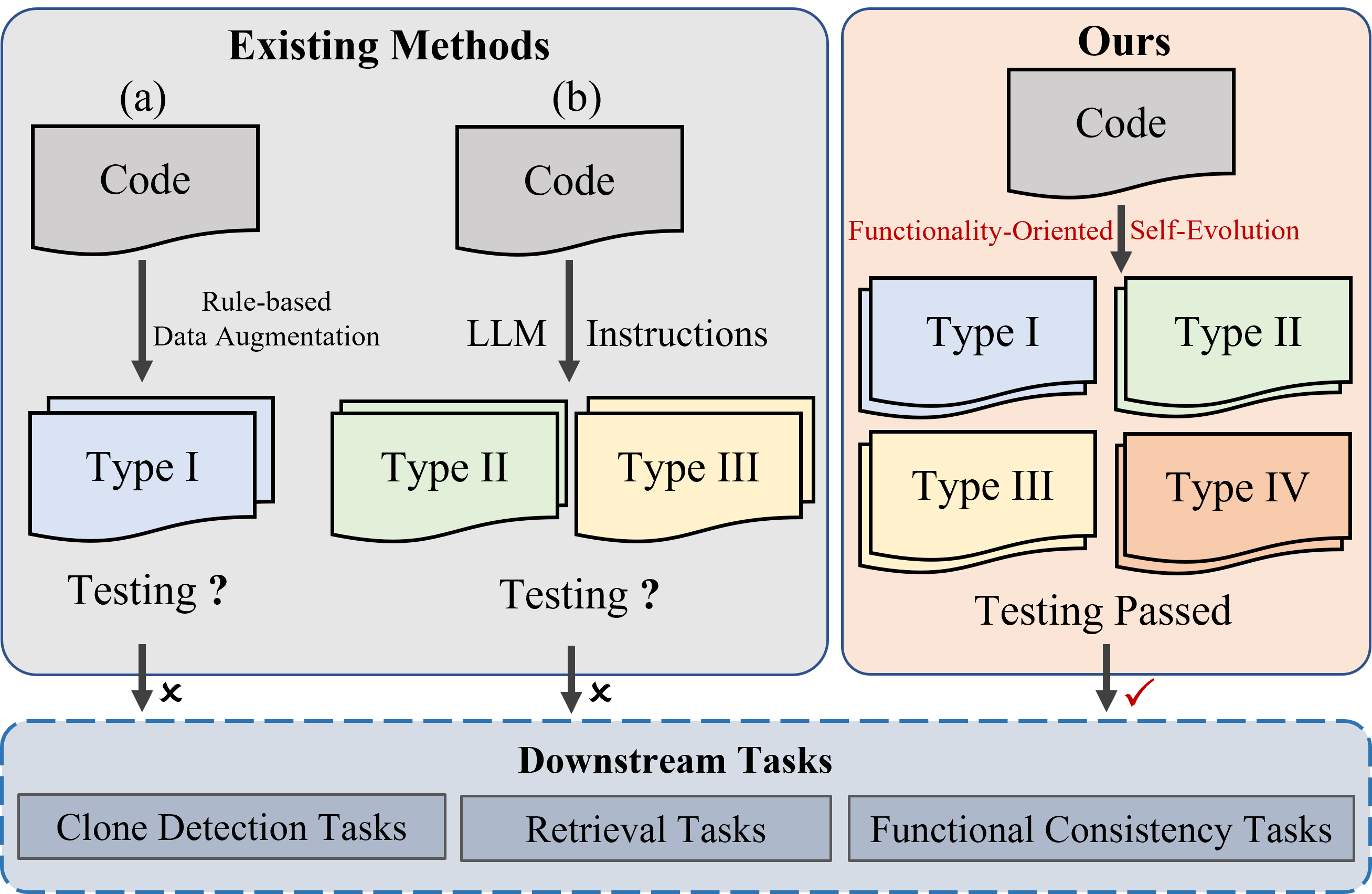}
  \caption{%
  \lzhmark{The comparison between our method and existing data synthesis methods.
  Existing methods fall into two categories: (a) rule-based augmentation and (b) instruction-based prompting via LLMs. While they produce partial variation types (e.g., Type I–III), they typically overlook challenging cases like Type IV and lack functionality validation. In contrast, our Functionality-Oriented Code Self-Evolution framework can cover all code scenarios, meeting the diversity and complexity requirements for downstream tasks.} \label{fig:methods_compare}
  }
\end{figure}

\lzhmark{To obtain more complex code datasets, existing data synthesis methods can be broadly divided into two categories, as illustrated in Figure~\ref{fig:methods_compare}: \textbf{(a)} rule-based data augmentation~\cite{karimi2021aeda,choi2024autoaugmentneedenhancingrulebased,fowler2018refactoring}, which generates limited variations (e.g., Type I samples via simple renaming or reordering of code)~\cite{liu2023contrabert}, and \textbf{(b)} instruction-based approaches using LLMs~\cite{li2024rewriting,cummins2024don}, such as EvoEval~\cite{evoeval} and WizardLM~\cite{xu2023wizardlm,luo2024wizardcoder}, which generate rewritten code (Type II) or significantly different code (Type III) by prompting LLMs to explore variations in both breadth (e.g., structural reformulations) and depth (e.g., reasoning steps or logic complexity).}
However, these approaches still fail to cover the full spectrum of functional consistency scenarios, particularly overlooking Type IV cases, and often lack effective mechanisms to verify the functional correctness of the generated code.

\lzhmark{To address these limitations, in this paper, we propose a novel data synthesis framework called Functionality-Oriented Code Self-Evolution, which systematically generates code variants across all four types (Table~\ref{tab:types}) from a single code instance.}
By evolving code along both syntactic and semantic dimensions, the framework produces a dataset that more accurately captures the complexities of real-world code scenarios.
Importantly, all generated code is accompanied by test cases to validate functional correctness, ensuring the dataset's reliability and utility for downstream tasks.
\lzhmark{\textit{
Unlike previous studies that often cover just part of the functional consistency scenarios, our approach handles all four types within a single framework.
}}

Finally, to evaluate the effectiveness of our approach, we conducted extensive experiments on three code-related tasks: code clone detection, code functional consistency identification, and code retrieval. Our preliminary results revealed that current datasets are insufficient for embedding models to achieve high-accuracy functional consistency identification, especially in complex scenarios. We then applied the evolved dataset to state-of-the-art embedding models, such as bge-en-icl~\cite{bge-en-icl}, using task alignment through in-context learning and instructions. Experimental results across all three tasks showed consistent performance improvements, even when applied across different datasets like HumanEval~\cite{chen2021codex}. These findings highlight the potential of our framework to generate diverse and semantically rich code samples, enabling embedding models to better capture the functional nuances of code.

Our main contributions can be summarized as follows:
\begin{itemize}
\item[$\bullet$] 
We conducted an empirical study on existing code embeddings and identified their limitations in effectively handling complex code scenarios.
\item[$\bullet$] 
We propose a novel data synthesis framework called Functionality-Oriented Code Self-Evolution for benchmarking functional consistency. With a single code sample, our approach generates diverse code variants that evolve across both syntactic and semantic dimensions.

\item[$\bullet$] 
We introduce two evolved datasets to benchmark the functional consistency of LLM-based code embeddings. Fine-tuning on these datasets improves downstream tasks, transferring insights from code functional consistency to code retrieval.

\end{itemize}

\section{Related Work}

\subsection{Code LLMs}
With the rise of LLMs demonstrating powerful text processing and generation capabilities, Code Large Language Models (Code LLMs) have also emerged as an increasingly prominent research topic~\cite{depalma2024exploring,fan2023large,qu2025review}.
\revision{A series of Code LLMs, such as CodeBert~\cite{codebert}, UniXcoder~\cite{guo-etal-2022-unixcoder}, CodeLlama~\cite{codellama}, Qwen-Coder~\cite{hui2024qwen2}, and Deepseek-Coder~\cite{deepseek-coder}, have shown promising performance in the field of code.
The popularity of Code LLMs stems from their ability to adapt to various downstream tasks without the need for additional processing, i.e., the potential for zero-shot usage in real-world scenarios. Recent work like CodeSage employs a two-stage large-scale pretraining strategy to improve multilingual code representation~\cite{zhang2024code}.
Despite these advances, most existing Code LLMs are designed to model surface-level code patterns or generate plausible code snippets, without a specific focus on capturing deeper functional consistency between code segments.}

\revision{\subsection{LLM Embeddings}
\noindent\textit{\textbf{Text Representation Learning.}}
Embedding models encode textual data (input tokens) into a latent space, where the output embeddings express the underlying semantics of the input~\cite{reimers2019sentence,KOVACEVIC2022117607}.
Benchmarks such as the Massive Text Embedding Benchmark (MTEB)~\cite{muennighoff2022mteb} provide standardized evaluation across diverse downstream tasks, establishing a unified leaderboard. The rise of applications like Retrieval-Augmented Generation (RAG) highlights the growing utility of high-quality embeddings. Rigorous benchmarking has driven the development of powerful, general-purpose models that perform effectively across multiple tasks and modalities.\\
\noindent\textit{\textbf{Code Representation Learning.}}
While text embeddings have been extensively studied~\cite{muennighoff2022mteb}, code embeddings remain relatively underexplored. Code embedding models specialize in encoding the semantic information (deep features) of code into dense vector representations~\cite{coderetriever,chen2019literature,TIAN2021114348,zhang2024code}, supporting downstream tasks such as code retrieval~\cite{gu2021cradle,huang-etal-2021-cosqa,khan2023xcodeeval} and code clone detection~\cite{fang2020functional,lu2021codexglue,abid2025measuring}.
State-of-the-art embedding models, including BGE-M3~\cite{BGE-M3} and bge-en-icl~\cite{bge-en-icl,bge-en-icl2}, enable multilingual, multi-granularity retrieval and leverage in-context learning.
Recent advances focus on code-native representations, such as FuzzPretrain, which incorporates dynamic execution insights, have enhanced semantic representation and robustness in code understanding~\cite{huang2023code}.}

\subsection{Code Similarity Measurement}
Code similarity measurement has long been a central problem in software engineering and program analysis~\cite{ragkhitwetsagul2018comparison,kartal2024automating}.
Over the past years, the similarity of code pairs has been measured from two dimensions: the syntax of the source code (program text) and the semantics of the source code (program functionality)~\cite{roy2007survey}. More recently, the task of predicting code semantic consistency has been formally framed as the problem of code clone detection~\cite{svajlenko2014bcb,poj104,SHENEAMER2018405,SUDHAMANI201963,nashaat2025enhanced,lu2021codexglue}. 
\revision{Pre-trained models such as CodeBert~\cite{codebert}, ContraBert~\cite{liu2023contrabert}, and CodeSAM~\cite{mathai2024codesam} have achieved notable progress by learning joint representations of code syntax and semantics.}
However, existing datasets often conflate semantically similar code examples with overlapping syntactic features, which reduces the robustness of approaches in identifying true functional consistency.
Different from those works, in this paper, we address the more challenging task of code functional consistency identification, which goes beyond code clone detection by disregarding syntactic information during prediction.

\begin{table*}[ht]
    \caption{
    The evaluated embedding models are categorized based on their model parameter sizes: Million-level, Billion-level, and models with unknown parameter sizes. ``Pooling'' shows how/where embeddings are obtained. Multiple [CLS] is the special embedding pooling method for BGE-M3.
    ``N/A'' indicates the information is not available publicly.
    } 
    \centering
    \resizebox{1.0\textwidth}{!}
    {
        \begin{tabular}{ l | c c c  c }
            \toprule
            \textbf{Models} & \textbf{Parameters} & \textbf{Dimensions} & \textbf{Max Length} & \textbf{Pooling}\\ 
            \hline
            CodeBert & 125 Million & 768 & 514 & [CLS] token \\
            UniXcoder & 126 Million & 768 & 1026 & [CLS] token \\
            BGE-M3 & 560 Million & 1024 & 8192 & Multiple [CLS]\\
            \hline
            CodeLlama & 34 Billion & 8192 & 16384 & Pooling \\
            DeepSeek-Coder & 33 Billion & 7168 & 16384 & Pooling \\
            \hline
            text-embedding-ada-002 & N/A & 1536 & 8191 & N/A \\ 
            text-embedding-3-large & N/A & 3072 & 8191 & N/A \\ 
            Embedding-2 & N/A & 1024 & N/A & N/A \\
            \bottomrule            
        \end{tabular}
    }
    \label{tab:embedding_models}
\end{table*}

\section{\lzhmark{Functionality-Oriented Code Self-Evolution}}
\lzhmark{In this work, we first evaluate existing code embedding models and identify their limitations in capturing functional consistency. Based on these findings, we propose Functionality-Oriented Code Self-Evolution, a framework that systematically synthesizes functionally diverse code variants from a single sample.}
%
%

\subsection{Embeddings for Code Functional Consistency}
We detail our preliminary experiments, focusing on code embedding pooling and evaluation for code functional consistency.
\revision{Specifically, code pairs are input into the embedding models to yield vector representations, and their functional consistency is assessed using cosine similarity, which is one of the most widely adopted metrics in the field of embeddings.}
As shown in Table~\ref{tab:types}, our target models include three classes based on model parameters: Million-level, Billion-level (Code LLMs), and unknown-level (using via API).

\noindent\textit{\textbf{Pooling.}}
To incorporate external knowledge, some code models utilize independent text/code embedding technologies (e.g., text-embedding-ada-002~\cite{ada-002}, text-embedding-3) to extract features without retraining. In contrast, LLMs such as ChatGPT and CodeLlama~\cite{codellama} rely on hidden states as embeddings when no additional embedding methods are used.
As shown in Table~\ref{tab:embedding_models}, for existing models accessed via API, the methods for extracting input features remain unspecified. However, for Code LLMs, we can explicitly apply post-processing techniques, such as pooling, to the hidden states (i.e., extracted features).
Generally, for Code LLMs with Transformer-based architectures, code embeddings are typically obtained in two ways: 
\textbf{(a)} Using the vector at the [CLS] position, as in BERT-based models (e.g., CodeBert~\cite{codebert} and UniXcoder~\cite{guo-etal-2022-unixcoder});
\textbf{(b)} Applying pooling (mean or max) to the vectors from the model's last layer, common in models like CodeLlama~\cite{codellama} and Deepseek-Coder~\cite{deepseek-coder}.

\lzhmark{
\subsection{Self-Evolution for Code Functional Consistency}}
\lzhmark{
As shown in Table~\ref{tab:types}, we categorize all possible code scenarios into four types based on syntactic (shallow) and semantic (deep) features to better evaluate the effectiveness of the embeddings.
Given a source code $c$, a code sample will be classified as a positive sample (Type I or Type II) if it shares deep features with $c$, indicating consistent functionality. If deep features are not shared, the sample is considered a negative sample (Type III or Type IV).
Furthermore, based on the syntactic dimension, positive samples are further classified as Type I (with shallow features similar to $c$) and Type II (without shallow features similar to $c$). Similarly, negative samples are categorized as Type III and Type IV.
\textit{However, existing code synthesis methods (as illustrated in Figure~\ref{fig:methods_compare}) aim to increase data diversity by performing surface-level augmentation or LLM-based rewriting and fail to generate the full spectrum of consistency types for functionality-oriented embeddings.}}

\lzhmark{\subsubsection{Problem Setting}}
\lzhmark{
To systematically generate all four functional consistency types outlined in Table~\ref{tab:types}, we propose a formalized data synthesis setup: given a code sample $c$, we define a set of self-evolution directions $\{x_i\}$ that reflect different levels of syntactic and semantic variation.
Each self-evolution direction $x_i$ is represented as a function of semantic and syntactic change:
\begin{equation}
x_i = f(\Delta_{\text{sem}}, \Delta_{\text{syn}})
\label{eq:xi-delta}
\end{equation}
where $\Delta_{\text{syn}}$ indicates the degree of syntactic modification (e.g., structure, control flow, naming), and $\Delta_{\text{sem}}$ indicates the degree of semantic deviation (e.g., changes in functionality or output behavior).
Applying $x_i$ to the source code $c$ yields an evolved variant $y_i$:
\begin{equation}
y_i = \text{Evolve}(c, x_i) = \text{Evolve}\left(c, f(\Delta_{\text{sem}}, \Delta_{\text{syn}})\right)
\label{eq:evolve-delta}
\end{equation}
Each $(c, x_i, y_i)$ triplet corresponds to one of the four canonical code pair types, based on combinations of $\Delta_{\text{sem}}$ and $\Delta_{\text{syn}}$.
}

\begin{figure*}[t]
    \centering
        \centering
        \includegraphics[width=0.85\textwidth]{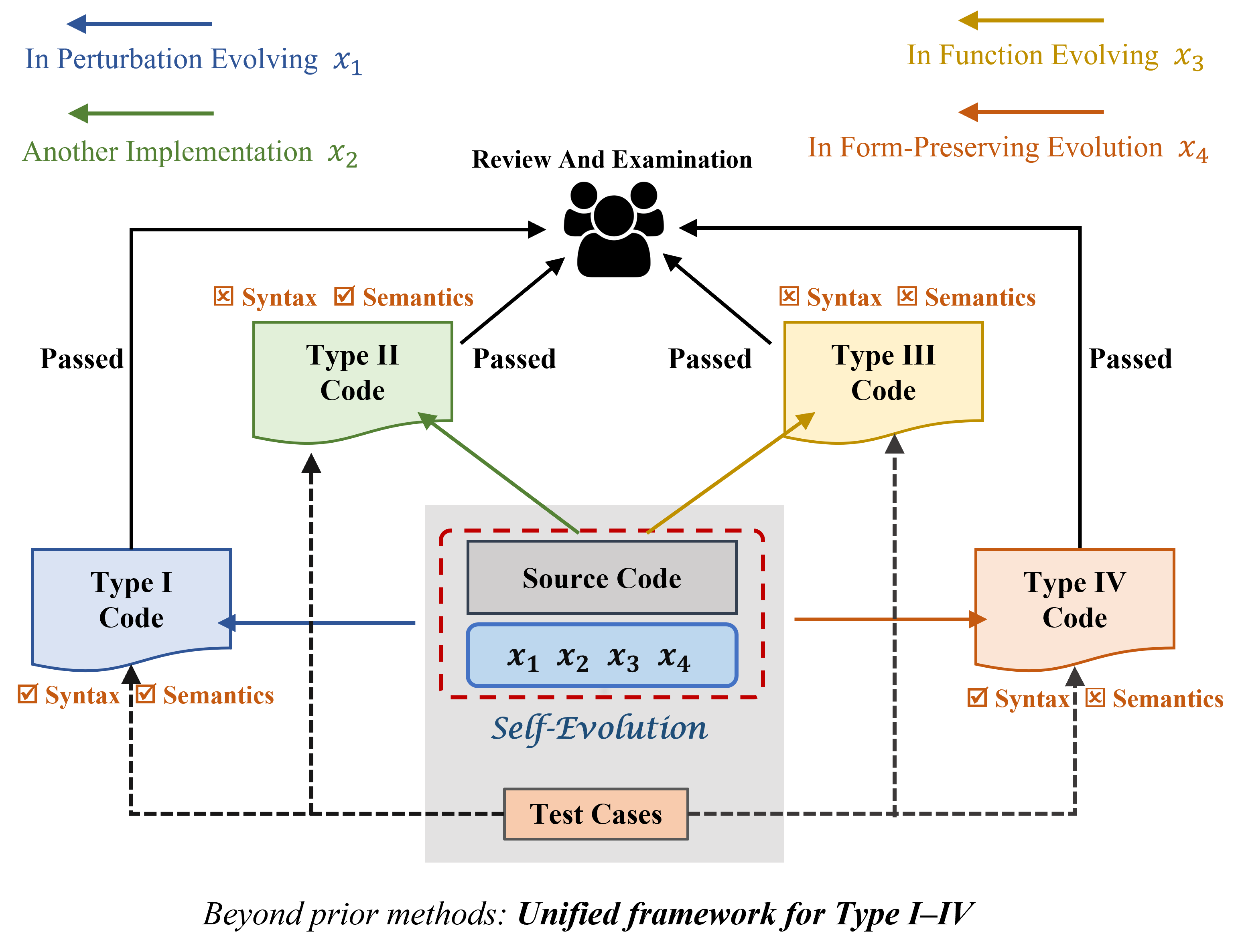}
    \caption{
    \lzhmark{Overview of the Functionality-Oriented Code Self-Evolution framework.
    Given a source code snippet, the framework uses only a single code instance to generate four code variations from syntactic and semantic aspects.}
}
    \label{fig:agu_arch_b}
    \vspace{-1em}
\end{figure*}

\subsubsection{Code Self-Evolution}
\label{datestes}
We now implement targeted evolution strategies to define four distinct evolution directions ($x_1$, $x_2$, $x_3$, $x_4$), each corresponding to a specific type of functional consistency.
These directions are characterized by their intended levels of syntactic and semantic variation, i.e., $\Delta_{\text{syn}}$ and $\Delta_{\text{sem}}$.
Specifically, inspired by data evolution paradigms in WizardLM~\cite{xu2023wizardlm} and WizardCoder~\cite{luo2024wizardcoder}, we proposed Functionality-Oriented Code Self-Evolution, where each direction $x_i$ guides the transformation of a source code 
$c$ into a corresponding variant $y_i$.
An overview of the Functionality-Oriented Code Self-Evolution variations is presented in Figure~\ref{fig:agu_arch_b}.

\revision{Despite the development of code augmentation techniques, there is still no unified and rigorous framework for validating the functional consistency of generated variants.
Prior works typically adopt either rule-based execution~\cite{liu2023contrabert, evoeval, song2024code} or metric-based heuristics~\cite{evtikhiev2023out}, and  often omit executability validation~\cite{luo2024wizardcoder}.}
\revision{In this work, to ensure consistent categorization of $y_i$, \textbf{we adopt a unified classification criterion that combines rule-based execution, metric-based filtering, and human-based verification.}} Specifically, $\Delta_{\text{sem}}$ is determined through automated test case execution: if $y_i$ passes the same test cases and produces outputs identical to $c$, it is considered semantically equivalent. This can be expressed as:
\begin{equation}
f_{\text{exec}}(c, y_i) = 
\begin{cases}
0, & \text{if } y_i \text{ passes } c\text{'s test cases;}\\
(0, 1], & \text{otherwise.}
\end{cases}
\end{equation}
where $f_{\text{exec}}(c, y_i)=0$ denotes semantic equivalence under test cases; otherwise, $\Delta_{\text{sem}} > 0$.

To assess $\Delta_{\text{syn}}$, we follow prior work~\cite{recode} and adopt CodeBLEU~\cite{ren2020codebleu} as the syntactic similarity metric between $y_i$ and $c$. 
\revision{Unlike traditional surface-level metrics such as BLEU, CodeBLEU captures n-gram overlap, abstract syntax tree structure, data-flow consistency, and syntax matching, which makes it more consistent with human-perceived code similarity~\cite{ren2020codebleu,evtikhiev2023out}. This richer representation enables CodeBLEU to more accurately reflect the structural fidelity of code transformations.
Since CodeBLEU can be affected by dataset characteristics (e.g., variations in source programming languages and coding conventions), we first manually filter the generated samples to match the definitions of the four functional code types. We then conduct a small-scale analysis on synthesized code samples to empirically determine an appropriate threshold $\theta$ for filtering generated variants. Further details are provided in Section~\ref{sec:experiments}.}
\revision{This can be expressed as: 
\begin{equation}
f_{\text{syn}}(c, y) = 
\begin{cases}
0, & \text{if } \text{CodeBLEU} \leq \theta ; \\
(0, 1], & \text{otherwise.}
\end{cases}
\end{equation} where $\Delta_{\text{syn}} = f_{\text{syn}}(c, y_i)$.}

\noindent\textit{\textbf{Type I.}}
\lzhmark{In perturbation evolving, LLMs perform random code perturbations on the input source code, including but not limited to changes in variable names, function names, comments, and dead code. This evolution corresponds to a case where $\Delta_{\text{sem}} = 0$ (i.e., functionality remains unchanged) while $\Delta_{\text{syn}} > 0$.
It ensures that the generated code maintains similar shallow features to the source, yet differs slightly in surface form.}

\noindent\textit{\textbf{Type II.}}
\lzhmark{To enable LLMs to generate code with the same functionality, we first have the LLM read the source code and analyze its functional intent.
Then, we instruct LLMs to produce diverse code samples that implement the same functionality, serving as another implementation.
This setting corresponds to $\Delta_{\text{sem}} = 0$ while $\Delta_{\text{syn}} \gg 0$, aiming to produce significantly different surface-level realizations of the same behavior.}

\noindent\textit{\textbf{Type III.}}
\lzhmark{In function evolving, LLMs first read the source code and analyze its functional intent. 
Based on the analysis, LLMs then evolve the code to obtain functionally diverse implementations with random preferences, constraints, multi-step reasoning, and by increasing the depth and breadth of the source code.
This evolution involves both semantic and syntactic divergence, i.e., $\Delta_{\text{sem}} > 0$ and $\Delta_{\text{syn}} \gg 0$, and serves to create functionally and structurally diverse negative examples.
}

\noindent\textit{\textbf{Type IV.}}
\lzhmark{The code evolution process is similar to Type III in Form-Preserving Functional Evolution. 
The core difference is that we instruct LLMs to generate code that preserves the original syntactic style (shallow features) of the source code.
This strategy targets $\Delta_{\text{sem}} > 0$ and $\Delta_{\text{syn}} \approx 0$, yielding deceptive samples that look similar to the original code but perform different functionality.}

It should be noted that code evolved by LLMs must pass corresponding test cases to verify its executability and functionality.
In this process, the above unified rule-based execution, metric-based filtering, and human-based verification combination allows us to operationally distinguish all four code types using a unified and reproducible process.
\revision{Manual review and examination are further introduced as a supplementary, post-processing quality assurance step after automatic filtering. For every 100 automatically accepted samples, 20\% are randomly selected for verification by two independent reviewers to confirm executability and adherence to the defined functional characteristics of the assigned code type.}
\revision{Note that if a sample fails either the rule-based or metric-based filter, it is discarded.
This strict criterion prioritizes functional correctness and avoids retaining borderline or ambiguous variants.}

\revision{The instructions for Functionality-Oriented Code Self-Evolution, which serve as prompts for the LLM to generate code variants, are summarized in Tables~\ref{tab:evol_instructions} and~\ref{tab:evol_typeIII}. 
Table~\ref{tab:evol_instructions} presents the main prompts for all types, while Table~\ref{tab:evol_typeIII} contains supplementary prompts specific to Type III, which direct diverse random evolutions.
These prompts are essential for driving the synthesis process and ensuring reproducibility. They have been designed to strictly follow the widely adopted \textit{Helpful, Honest, and Harmless (HHH)} principle to prevent the generation of deceptive or harmful code, and manual verification is employed to filter out any potentially unsafe or misleading outputs.
Full contents are provided in~\ref{sec:instru1}.}

\section{Experiments\label{sec:experiments}}
We evaluate code embeddings and identify limitations in current datasets in Section~\ref{sec:limitations}, and introduce Functionality-Oriented Code Self-Evolution, which is tested for effectiveness and generalization across downstream tasks on POJ-104 and HumanEval in Sections~\ref{code evolution}–\ref{sec:general}.

\subsection{Evaluation of Code Embeddings: Revealing Dataset Limitations} \label{sec:limitations}
\lzhmark{To assess whether current code embedding models effectively capture functional consistency, we evaluate both vanilla and fine-tuned versions on the CodexGLUE benchmarks (BigCloneBench~\cite{svajlenko2014bcb,wang2020bcb} and POJ-104~\cite{poj104}), using cosine similarity as the evaluation metric.}

\begin{table*}[ht]
    \caption{
    Vanilla embedding model performance on the Big Clone Bench dataset under different thresholds, revealing limited capability in capturing code functional consistency.
    } 
    \centering
    \tabcolsep=0.27cm
    {
        \resizebox{0.95\textwidth}{!}
        {\begin{tabular}{ l | l  l  l |  l  l  l c }
            \toprule
            & \multicolumn{3}{c}{\textbf{Base Threshold-0.5}} & \multicolumn{4}{c}{\textbf{Best Threshold}}\\
            \hline
            {\textbf{Models}}  & \textbf{Precision} & \textbf{Recall} & \textbf{F1} & \textbf{Precision} & \textbf{Recall} & \textbf{F1}  & \textbf{Threshold}\\ 
            \hline
            CodeBert \cite{codebert}
            & 0.132 & 0.980 & 0.233 & 0.137 & 0.999 & 0.241 & 0.4
            \\
            UniXcoder \cite{guo-etal-2022-unixcoder} 
            & 0.102 & 0.423 & 0.164 & 0.137 & 0.996 & 0.241 & 0.3
            \\
            BGE-M3 \cite{BGE-M3}
            & 0.158 & 0.973 & 0.273 & 0.316 & 0.630 & 0.421 & 0.6 
            \\   
            \hline
            CodeLlama \cite{codellama}
            & 0.137 & 0.986 & 0.241 & 0.194 & 0.449 & 0.271 & 0.8
            \\
            DeepSeek-Coder \cite{deepseek-coder} 
            & 0.138 & 0.994 & 0.242 & 0.139 & 0.958 & 0.243 & 0.6
            \\
            \hline
            text-embedding-ada-002
            & 0.137 & 1.0 & 0.241 & 0.580 & 0.359  & 0.444 & 0.8
            \\  
            text-embedding-3-large
            & 0.670 & 0.205 & 0.314 & 0.440  & 0.464 & \textbf{0.452} & 0.4 
            \\
            Embedding-2 & 0.344 &0.596 & \textbf{0.436} & 0.344 &0.596 & 0.436 & 0.5
             \\            
            \bottomrule            
        \end{tabular}
    }}
    \label{tab:compare_bcb}
    \vspace{-0.5em}
\end{table*}

\begin{table}[!t]
    \caption{
    Vanilla embedding model performance on the POJ-104 dataset, revealing limited capability in capturing code functional consistency.
    } 
    \centering
    \tabcolsep=0.27cm
    \resizebox{0.5\textwidth}{!}{
        \begin{tabular}{ l |  c }
            \toprule
            {\textbf{Models}}  & \textbf{MAP@R(\%)}  \\ 
            \hline
            CodeBert \cite{codebert} & 2.55 \\
            UniXcoder \cite{guo-etal-2022-unixcoder} & 29.9\\
            BGE-M3 \cite{BGE-M3} & 42.85
            \\  
            \hline
            CodeLlama \cite{codellama} & 28.6
            \\
            DeepSeek-Coder \cite{deepseek-coder} & 42.9
            \\
            \hline
            text-embedding-ada-002 & \textbf{73.25} \\  
            text-embedding-3-large & 71.75 \\   
            Embedding-2 & 62.25 \\          
            \bottomrule            
        \end{tabular}
    }
    \label{tab:compare_poj}
\end{table}

\begin{table}[!t]
    \caption{
    Performance of embedding models (w/ \& w/o fine-tuning) on Big Clone Bench and POJ-104, highlighting that existing datasets lack complexity and fail to reflect diverse code scenarios.
    } 
    \centering
    \tabcolsep=0.27cm
    \resizebox{0.5\textwidth}{!}
    {
        \begin{tabular}{ l |  c }
            \toprule
            \textbf{Big Clone Bench}  & \textbf{F1}  \\ 
            \hline
            CodeBert \cite{codebert} & 0.233 (\rt{+0.708})\\
            CodeBert + Finetune \cite{codellama} & 0.941 \\
            UniXcoder \cite{guo-etal-2022-unixcoder} & 0.164 (\rt{+0.773})\\
            UniXcoder + Finetune \cite{guo-etal-2022-unixcoder} & 0.937 \\
            \bottomrule  
            \textbf{POJ-104}  & \textbf{MAP@R(\%)} \\ 
            \hline
            CodeBert \cite{codebert} & 2.55 (\rt{+80.12})\\
            CodeBert + Finetune \cite{codellama} & 82.67 \\
            UniXcoder \cite{guo-etal-2022-unixcoder} & 29.9 (\rt{+60.74})\\
            UniXcoder + Finetune \cite{guo-etal-2022-unixcoder} & 90.64 \\
            BGE-M3 \cite{BGE-M3} & 42.85 (\rt{+28.9}) \\     
            BGE-M3 + Finetune \cite{BGE-M3} & 71.75 \\    
            \bottomrule            
        \end{tabular}
    }
    \label{tab:compare_finetune}
    \vspace{-1em}
\end{table}

\noindent\textit{\textbf{Datasets.}}
\textbf{(a)} BigCloneBench (BCB) dataset
is the largest dataset for clone detection, including projects from 25,000 systems and ten functionalities.
For BCB, given two codes as the input, the task is to do binary classification (0/1), where 1 stands for semantic equivalence and 0 for others.
Models are evaluated by F1 score, precision, and recall, with F1 as the main metric.
\textbf{(b)} POJ-104 dataset, contains 104 programming tasks and 500 different codes for each task. 
Models are evaluated by MAP@R score, which is the mean of average precision scores for predicting the R most similar samples for a given query. 
R is the number of other codes in the same class, i.e., 499.

\noindent\textit{\textbf{Models.}}
We evaluate four popular open-source embedding models for code tasks: three Million-level models (CodeBert~\cite{codebert}, UniXcoder~\cite{guo-etal-2022-unixcoder}, BGE-M3~\cite{BGE-M3}) and two Billion-level LLMs (CodeLlama~\cite{codellama} and Deepseek-Coder~\cite{deepseek-coder}). The largest and best-performing versions, Codellama-34B and Deepseek-Coder-33B, are used for a comprehensive evaluation of embeddings.
For BERT-based models, we follow the CodeXGLUE setup and use the [CLS] token for embeddings. For LLMs, we apply mean pooling to the hidden states. Additionally, we evaluate state-of-the-art embedding models for text/code tasks, including three black-box models: text-embedding-ada-002, text-embedding-3 from OpenAI, and Embedding-2 from MetaGLM.
%
For black-box models, we specify versions via API.

\noindent\textit{\textbf{Results.}}
The experimental results of the vanilla embedding models on the BCB and POJ-104 datasets are presented in Tables~\ref{tab:compare_bcb} and~\ref{tab:compare_poj}, respectively. The results of the fine-tuned models are presented in Table~\ref{tab:compare_finetune}.
We adopt a base detection threshold of 0.5 for BCB dataset, consistent with prior work~\cite{svajlenko2014bcb,wang2020bcb,lu2021codexglue}. 
%
After analyzing the results, we find that:
Vanilla embedding models, including LLMs, lead to sub-optimal performance and fail to capture code semantics effectively.
For API-based models, the best F1 scores on the BCB dataset show little variation across models, but on the POJ-104 dataset, performance differences are more pronounced.
However, after fine-tuning, all three Million-level models show significant improvements. For instance, CodeBert’s accuracy on POJ-104 increased by nearly 40$\times$ after fine-tuning, surpassing 90\% accuracy.

These findings suggest that datasets like POJ-104 have reached a complexity bottleneck, being overly simplistic and lacking challenging, comprehensive code scenarios. 
As a result, \textbf{existing datasets fail to adequately evaluate the true effectiveness of embedding models in capturing of code functional consistency, highlighting the need for more diverse and complex code scenarios to evaluate true model performance.}
Consequently, we propose evolving the source code to cover a broader range of possible code scenarios, which will better challenge embedding models and provide a more accurate evaluation.



\subsection{Construction and Evaluation of the Code Self-Evolution} \label{code evolution}

Through the proposed Functionality-Oriented Code Self-Evolution, we construct a code evolution dataset that expands the dataset in terms of both shallow features (syntax) and deep features (semantics). 
We demonstrate the construction process using POJ-104 as a representative example, given its task diversity and prevalence in code understanding literature. Notably, this framework is not tailored to POJ-104. The same procedures and evolution strategies are applied to other benchmarks such as HumanEval, demonstrating its adaptability across datasets (see Section~\ref{sec:general}).

\noindent\textit{\textbf{Experimental Setup.}}
To ensure the diversity of evolved code, we utilized a mixture of \textit{gpt-3.5-turbo-0125} and \textit{gpt-4-turbo-2024-04-09} models as the LLMs for evolving the source code. Generation was executed through the OpenAI API using default hyperparameters: temperature = 1.0, top\_p = 0.95, and max\_tokens = 2048. 
\revision{Code generation primarily relies on LLMs, with costs proportional to API calls and the success rate of generating valid code samples. Each API call generates one candidate sample based on the corresponding instruction. Since each API call is an independent process, our approach is highly parallelizable. Consequently, both computational and monetary costs exhibit linear scalability with the number of generated samples.}
This process was performed on a laptop equipped with an AMD Ryzen 7 8845HS CPU (3.80 GHz) and 24 GB of RAM. 

\noindent\textit{\textbf{\ourdataset{}.}}
First, we randomly selected a code sample from each programming task in POJ-104 and provided it to ChatGPT. We then prompted the model to summarize the intended functionality of code and generate unified test cases for the entire task.
By executing the unified test case, we excluded unusable code samples from the POJ-104 dataset, e.g., those with missing functions, incorrect return types, or other compilation issues.
Rather than generating separate test cases for each source code sample, we used a unified programming example per task as a shared test case for all code variants within the same task.

Second, among the four types, it is straightforward to obtain Type II code.
We provided ChatGPT with the task description of the source code and asked it to generate another implementation. 
We then selected the Type II codes that passed the test cases.
For the remaining three code versions evolved by LLMs, taking the evolution route of Type I code as an example, we provided ChatGPT with the source code to be evolved and corresponding evolution instructions. 
ChatGPT generated a new population of code through perturbation evolution. 
Since the semantics of the code did not change, we could select available evolved Type I code from a series of candidates by running the corresponding test cases.
In function evolving and form-preserving functional evolution, we will obtain the populations of Type III and Type IV code candidates evolved by ChatGPT, respectively. 
To ensure that Type III and Type IV codes are negative samples of the source code, performing different functionalities with the source code, we leveraged the advantage of using the test cases without additional information.
Specifically, we input the test cases into the evolved code candidates. 
If the program successfully executes and produces outputs different from those of the test cases, we can obtain the available evolved Type III and Type IV codes.

\revision{Finally, to further ensure the completeness and usability of \ourdataset{}, we introduced two additional verification steps: (i) metric-based filtering using CodeBLEU~\cite{ren2020codebleu,lu2021codexglue} to re-evaluate the syntactic similarity of code, and (ii) human-based verification through manual review and examination.}
Specifically, evolved codes were evaluated using CodeBLEU to compare with the source code and subjected to a threshold filter. 
\revision{As described in Section~\ref{datestes}, our metric-based filtering stage selects syntactically diverse code variants using a CodeBLEU threshold $\theta$. To empirically determine $\theta$, we conducted a controlled study on 100 code variants sampled across all evolution types.
Each variant was independently labeled by two human annotators as structurally similar or dissimilar to its source code. We then evaluated candidate thresholds from 0.2 to 0.6, treating CodeBLEU as a binary classifier against the human annotations, and performance was measured via F1-score. The highest alignment (F1 = 0.94) was achieved at $\theta$ = 0.4, indicating the best alignment with human judgment. Figure 1 further illustrates this finding through representative examples annotated with their CodeBLEU scores (Type I = 0.7059, Type II = 0.2017, Type III = 0.1234, Type IV = 0.4455). These results provide converging quantitative and visual evidence that $\theta = 0.40$ is an effective criterion for metric-based filtering.}

Additionally, evolved codes were randomly selected for manual inspection to confirm the correctness of their types.
With the above experimental setup and steps, we divided the evolved dataset into training and testing sets in a ratio of 3:1.
The details of the evolved code dataset are shown in Table~\ref{tab:evolved_dataset}.

\begin{table}[!t]
    \caption{\lzhmark{Our evolved code dataset, \ourdataset{}, contains four different code types with more challenging and complete pairwise scenarios than POJ-104.}}
    \centering
    \tabcolsep=0.27cm
    \resizebox{0.45\textwidth}{!}
    {
        \begin{tabular}{ c l | c c }
            \toprule
            \multicolumn{2}{c}{\textbf{Code Types}} & \textbf{Train} & \textbf{Test} \\ 
            \hline
            \multirow{2}{*}{Positive} & Type I & 1546 & 538\\ 
            \cline{3-4}
            & Type II  & 1518 & 533\\
            \hline
            \multirow{2}{*}{Negative} & Type III & 1513 & 548\\
            \cline{3-4}
            & Type IV  & 1626 & 813 \\
            \bottomrule       
        \end{tabular}
    }
    \label{tab:evolved_dataset}
\end{table}

\begin{table}[!t]
    \caption{
    \lzhmark{Dataset Effectiveness Comparison on the BGE-M3 in Terms of F1 Score. Our evolved dataset leads to higher F1, validating its superior data quality.}}
    \centering
        \resizebox{0.6\textwidth}{!}
        {
        \begin{tabular}{ l | c | c | c }
            \toprule
             & \textbf{M3} & \textbf{M3\_poj} & \gt{\textbf{M3\_evl}} \\ 
            \hline
            \textbf{Type I} & 1.0 & 1.0 & \gt{1.0}\\ 
            \hline
            \textbf{Type II}  & 1.0 & 1.0 & \gt{1.0}\\
            \hline
            \textbf{Type III} & 0.0383 \rt{(+0.8065)} & 0.2956 \rt{(+0.5492)} & \gt{0.8448}\\
            \hline
            \textbf{Type IV}  & 0.0025 \rt{(+0.7404)} & 0.1488 \rt{(0.5941)} & \gt{0.7429}\\ 
            \hline
            \textbf{Mean} & 0.4498 \rt{(+0.4334)} & 0.5567 \rt{(+0.3265)} & \gt{0.8832} \\
            \bottomrule       
        \end{tabular}
    }
    \label{tab:test_evolved_dataset}
\end{table}

\noindent\textit{\textbf{Effectiveness of \ourdataset{}.}}
Considering the landmark work in text embeddings~\cite{muennighoff2022mteb}, which suggests that embedding models do not reach the billion-level in terms of model parameters, we selected three Million-level models to validate the effectiveness of our \ourdataset{}.
The code detection threshold used is the best threshold for BGE-M3 on BCB dataset, 0.6 as shown in Table~\ref{tab:compare_bcb}.
Similarly, the thresholds for the two BERT models are 0.4 and 0.3, respectively.
The effectiveness of our dataset was evaluated via F1 score comparisons for M3, as shown in Table~\ref{tab:test_evolved_dataset}.
Note that the experimental results of the BERT-based models show trends consistent with those of the M3 model, further confirming the effectiveness of the proposed dataset. For detailed results, please refer to the~\ref{appendix-bert}.
M3 represents the vanilla BGE-M3 model~\cite{BGE-M3}, M3\_{poj} represents M3 trained on the POJ-104 dataset, and M3\_{evl} represents M3 trained on our proposed evolved dataset.
Similarly, BERT-based models follow the same naming rules.
\ding{182} M3's code embeddings tend to assign higher similarity to all codes, i.e., all positive codes are correctly classified, while almost all negative codes are incorrectly classified. This strongly indicates M3's unreliability in capturing code semantics.
\ding{183} M3\_{poj} shows marginal improvement in handling negative samples over M3, but its performance remains insufficient. For instance, the F1 score for Type III is below 0.3 and for Type IV it is around 1/8, highlighting POJ-104’s limited code diversity and inability to cover various code pairing scenarios.
\ding{184} M3\_{evl} achieved significantly higher performance with an overall accuracy of 0.88, 
outperforming other evaluated models.
\textit{It improved accuracy in Type IV scenarios by nearly 300$\times$ and 5$\times$ compared to M3 and M3\_{poj}, respectively, demonstrating that our \ourdataset{} dataset, with greater code diversity, surpasses POJ-104 in code functional consistency analysis.}

\subsection{Effectiveness Validation of Code Self-Evolution on Downstream Tasks}\label{sec:downstream}
\revision{To validate the practical effectiveness of functionality-oriented code self-evolution, we evaluate its impact across three representative downstream tasks: code clone detection, code functional consistency identification, and code retrieval, using task-specific evaluation metrics for each task.}
\revision{We employ the state-of-the-art embedding models~\cite{muennighoff2022mteb}, bge-en-icl~\cite{bge-en-icl,bge-en-icl2}, with task alignment strategies (in-context learning and instruction) where applicable.}
Detailed instructions for task alignment are provided in~\ref{sec:instru2}.

\noindent\textit{\textbf{Clone Detection Task.}}
\revision{Table~\ref{tab:code_clone} shows that vanilla embeddings from the bge-en-icl model already perform well on POJ-104 (MAP@R = 90.76\%), suggesting the dataset alone is no longer sufficient for a comprehensive evaluation of code embedding performance.}
\revision{
Following task alignment on the POJ-104 dataset, the performance improved as expected. Incorporating our POJ-Evl dataset as an additional enhancement further increases the detection accuracy to 95.60\%, indicating that our method yields more diverse and comprehensive code variants, thereby continuing to improve code representation capabilities even at high accuracy levels.}

\begin{table}[!h]\small
    \caption{
    \revision{Effectiveness comparison of bge-en-icl on the code clone detection task in terms of MAP@R, where \textit{Vanilla Embeddings} denotes the vanilla model, \textit{Task Alignment} denotes task alignment on POJ-104, and \textit{POJ-Evl} denotes Task Alignment further enhanced on our evolved dataset. Reported as mean $±$ standard deviation across five runs. * indicates statistical significance ($p < 0.05$) vs. Task Alignment (paired t-test).}}
    \centering
    \tabcolsep=0.27cm
    \resizebox{0.5\textwidth}{!}
    {
        \begin{tabular}{ l |  c }
            \toprule 
            \textbf{Code Clone Task}  & \textbf{MAP@R(\%)} \\ 
            \hline
            Vanilla Embeddings & 90.76 \scriptsize{\rt{(+4.84)}}\\
            + Task Alignment & 94.30 $\pm$ 0.55 \scriptsize{\rt{(+1.30)}}\\
             + POJ-Evl & \gt{95.60 $\pm$ 0.18} $^{*}$ \\
            \bottomrule            
        \end{tabular}
    }
    \label{tab:code_clone}
\end{table}

\begin{figure}[t]
  \centering
  \includegraphics[width=0.6\textwidth]{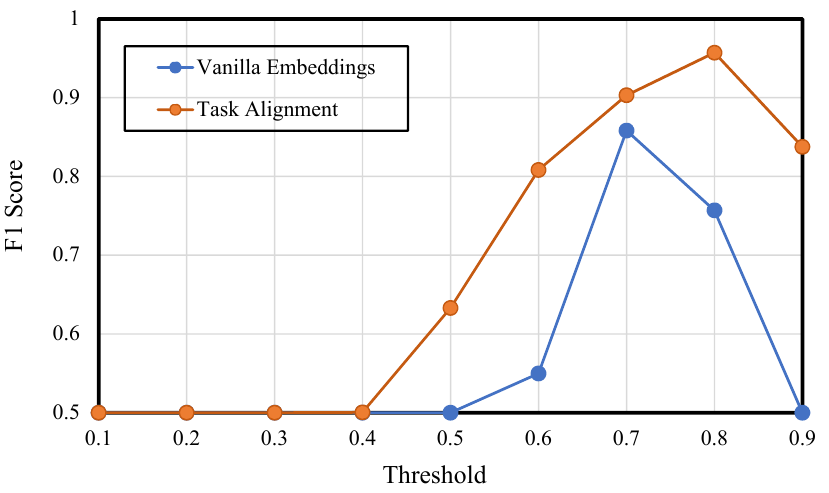}
  \caption{
      \revision{Comparison results for the functional consistency identification task (thresholds 0.1-0.9, step = 0.1; lower thresholds classify more code pairs as functionally consistent, while higher thresholds impose stricter requirements), showing consistent gains from our evolved dataset across all settings.}
  }
  \label{fig:threshold_consistency}
\end{figure}

\noindent\textit{\textbf{Functional Consistency Task.}}
As shown in Table~\ref{tab:code consistency}, vanilla embeddings assign high similarity scores to functionally dissimilar code (Types III/IV).
As expected, this issue was effectively addressed after task alignment using our POJ-Evl dataset, leading to significant improvements for negative types. This demonstrates that our data synthesis framework is capable of tackling such a highly challenging task.
\revision{To further assess robustness, we report the average accuracy across thresholds ranging from 0.1 to 0.9 (step = 0.1).} As illustrated in Figure~\ref{fig:threshold_consistency}, embeddings aligned with POJ-Evl consistently outperformed vanilla ones across all thresholds, reinforcing the method's stability and its ability to improve semantic discrimination regardless of threshold setting.

\begin{table}[H]
        \caption{\revision{Effectiveness comparison of bge-en-icl on the code functional consistency task in terms of F1 score, where \textit{Vanilla Embeddings} denotes the vanilla model, and \textit{POJ-Evl} denotes Task Alignment further enhanced on our evolved dataset. Reported as mean $±$ standard deviation across five runs. ** indicates statistical significance ($p < 0.01$) vs. vanilla embedding (paired t-test).}}
    
    \centering
    \resizebox{0.72\textwidth}{!}
        {
        \begin{tabular}{ c | c | c }
            \toprule
             \textbf{Functional Consistency}& \textbf{Vanilla Embeddings} & \textbf{+ POJ-Evl} \\ 
            \hline
            \textbf{Type I} & 1.0 & \gt{1.0 $\pm$ 0}\\ 
            \hline
            \textbf{Type II}  & 1.0 & \gt{1.0 $\pm$ 0}\\
            \hline
            \textbf{Type III} & 0.1703 \scriptsize{\rt{(+0.5225)}} & \gt{0.6928 $\pm$ 0.0221 $^{**}$}\\
            \hline
            \textbf{Type IV}  & 0.0291 \scriptsize{\rt{(+0.5304)}} & \gt{0.5595 $\pm$ 0.0107 $^{**}$}\\ 
            \hline
            \textbf{Mean} & 0.5426 \scriptsize{\rt{(+0.2705)}} & \gt{0.8131 $\pm$ 0.0055 $^{**}$} \\
            \bottomrule       
        \end{tabular}
    }
    \label{tab:code consistency}
\end{table}

\noindent\textit{\textbf{Retrieval Task.}}
We used the Code-Code retrieval task datasets from xCodeEval~\cite{khan2023xcodeeval}, a large-scale code-related task dataset. We computed the top-k retrieval accuracy (Acc@k\%) and reported the experimental results, as shown in Table~\ref{tab:code retrieval}, where k = 50.
Despite the small scale of the POJ-Evl dataset, it improved the semantic representation of code embeddings and enhanced task performance after alignment, demonstrating its effectiveness in boosting retrieval accuracy.

\begin{table}[H]
    \caption{
    \revision{Effectiveness comparison of bge-en-icl on the code retrieval task in terms of Acc@k, where \textit{Vanilla Embeddings} denotes the vanilla model, \textit{Task Alignment} denotes task alignment on xCodeEval, and \textit{POJ-Evl} denotes Task Alignment further enhanced on our evolved dataset. Reported as mean $±$ standard deviation across five runs. * indicates statistical significance ($p < 0.05$) vs. Task Alignment (paired t-test).}
    }
    \centering
    \tabcolsep=0.27cm
    \resizebox{0.52\textwidth}{!}
    {
        \begin{tabular}{ l |  c }
            \toprule
            \textbf{Code Retrieval Task}  & \textbf{Acc@k(\%)}  \\ 
            \hline
            Vanilla Embeddings & 32.45 \scriptsize{\rt{(+10.91)}}\\
             + Task Alignment & 42.25 $\pm$ 1.23 \scriptsize{\rt{(+1.11)}}\\
             + POJ-Evl & \gt{43.36 $\pm$ 1.04} $^{*}$ \\
            \bottomrule            
        \end{tabular}
    }
    \label{tab:code retrieval}
\end{table}

Overall, Functionality-Oriented Code Self-Evolution consistently improves model performance across diverse tasks. By addressing existing challenges in code representation, the proposed framework fills a critical gap, offering a powerful solution that is crucial for advancing the performance of code-related tasks in real-world applications.

To further illustrate the advantage of the proposed method, we compared Functionality-Oriented Code Self-Evolution with the rule-based augmentation method [16] on the code functional consistency task.
Table~\ref{tab:rule_based_vs_poj_evl} shows that our method consistently outperforms both the Vanilla embeddings and the rule-based augmentation across all categories. Notably, for the challenging Type III and Type IV, POJ-Evl achieves substantial improvements (0.6890 and 0.5443, respectively) compared to the modest gains from rule-based augmentation, demonstrating robust enhancement of dataset diversity and complexity.

\subsection{Generalization of Functionality-Oriented Code Self-Evolution} \label{sec:general}

Functionality-Oriented Code Self-Evolution is proposed to enrich the semantic depth and functional diversity of code datasets, with the overarching goal of demonstrating its generalization and adaptability across a wide array of code datasets.
We conducted a series of studies on the widely recognized and challenging HumanEval dataset~\cite{chen2021codex}, which serves as an independent benchmark, distinct from POJ-104.
HumanEval presents a more complex testing ground compared to POJ-104, comprising 164 algorithmically intricate Python programming tasks. These tasks are significantly more diverse and complex, making the dataset an ideal candidate to assess the scalability and robustness of our framework. We applied our evolution framework to generate evolved samples across all four code types (Type I–IV), resulting in the enhanced dataset, HumanEval-Evl, which notably boosts its evaluation potential. 
\revision{Consistent with our setting for the POJ-Evl dataset, we set the CodeBLEU threshold for HumanEval-Evl to $\theta = 0.5$, treating scores below this value as structurally dissimilar to the source code.}

\begin{table*}[ht]
    \caption{
    \lzhmark{Compared to HumanEval, HumanEval-Evl adds four distinct and challenging code types supporting functional consistency analysis.}}
    \centering
    \tabcolsep=0.27cm
    \resizebox{0.5\textwidth}{!}
    {
        \begin{tabular}{ c l | c c }
            \toprule
            \multicolumn{2}{c}{\textbf{Code Types}} & \textbf{Train} & \textbf{Test} \\ 
            \hline
            \multirow{2}{*}{Positive} & Type I & 817 & 120\\ 
            \cline{3-4}
            & Type II  & 632 & 110\\
            \hline
            \multirow{2}{*}{Negative} & Type III & 831 & 120\\
            \cline{3-4}
            & Type IV  & 654 & 120 \\
            \bottomrule       
        \end{tabular}
    }
    \label{tab:evolved_humanevl}
\end{table*}
We evaluated the effectiveness of HumanEval-Evl by applying it to the code functional consistency task described in Section~\ref{sec:downstream}. 

\begin{figure}[t]
  \centering
  \includegraphics[width=0.6\textwidth]{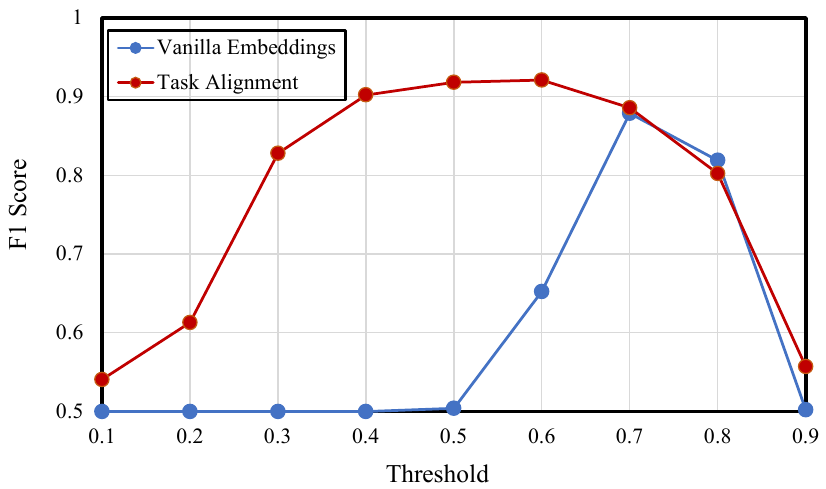}
  \caption{
    \revision{Comparison results for the functional consistency identification task (thresholds 0.1-0.9, step = 0.1; lower thresholds classify more code pairs as functionally consistent, while higher thresholds impose stricter requirements), showing consistent gains from our evolved dataset across all settings.}
  }
  \label{fig:threshold_consistency_he}
\end{figure}

As shown in Table~\ref{tab:code consistency-human_eval}, the results demonstrate that the proposed framework generalizes effectively to the HumanEval dataset.
Across all four code types, it consistently improves F1 scores, with especially substantial gains observed on functionally divergent negative samples---Type III (+0.4866) and Type IV (+0.5833).
We also assessed the performance of code embeddings by evaluating average accuracy across thresholds ranging from 0.1 to 0.9, as depicted in Figure~\ref{fig:threshold_consistency_he}. The results unequivocally show that embeddings aligned with our evolved dataset consistently outperform others across various thresholds, underscoring the superior quality and effectiveness of the code representations derived from our method.

\begin{table}[H]
    \caption{\revision{Effectiveness comparison of bge-en-icl on the code functional consistency task in terms of F1 score, where \textit{Vanilla Embeddings} denotes the vanilla model, and \textit{HumanEval-Evl} denotes Task Alignment further enhanced on our evolved dataset.}} 
    \centering
    \resizebox{0.72\textwidth}{!}
        {
        \begin{tabular}{ c | c | c }
            \toprule
             \textbf{Functional Consistency}& \textbf{Vanilla Embeddings} & \textbf{+ HumanEval-Evl} \\ 
            \hline
            \textbf{Type I} & 0.8917 \rt{(+0.0023)} & \gt{0.8940}\\ 
            \hline
            \textbf{Type II}  & 0.9182 \rt{(+0.0015)} & \gt{0.9197}\\
            \hline
            \textbf{Type III} & 0.4083 \rt{(+0.4866)} & \gt{0.8949}\\
            \hline
            \textbf{Type IV}  & 0.3917 \rt{(+0.5833)} & \gt{0.9750}\\ 
            \hline
            \textbf{Mean} & 0.6525 \rt{(+0.2684)} & \gt{0.9209} \\
            \bottomrule       
        \end{tabular}
    }
    \label{tab:code consistency-human_eval}
\end{table}

These findings decisively confirm that our proposed framework not only enhances the diversity and complexity of existing datasets but also elevates their performance on highly challenging and representative benchmarks like HumanEval. This substantiates the generalization of our approach, demonstrating its broad applicability to various code datasets, regardless of their origin or complexity.

\subsection{\revision{Comparison with Alternative Augmentation Methods}}
\revision{Most existing data augmentation methods aim to increase sample quantity~\cite{karimi2021aeda,choi2024autoaugmentneedenhancingrulebased,fowler2018refactoring,liu2023contrabert}, while our framework addresses the previously overlooked limitation of code datasets in functional diversity and semantic complexity.
To substantiate this, we compared Functionality-Oriented Code Self-Evolution with the rule-based augmentation methods~\cite{fowler2018refactoring} on the code functional consistency task.}

\begin{table}[H]
    \caption{\revision{Effectiveness Comparison of bge-en-icl on the Code Functional Consistency Task in Terms of F1 Score Across Different Augmentation Strategies, Highlighting Vanilla Embeddings, Rule-based Augmentation, and our Functionality-Oriented Code Self-Evolution.
    }}
    \centering
    \resizebox{0.75\textwidth}{!}
        {
        \begin{tabular}{ c | c | c | c}
            \toprule
             \textbf{Functional Consistency}& \textbf{Vanilla Embeddings} & \textbf{Rule-based Augmentation} & \textbf{Ours} \\ 
            \hline
            \textbf{Type I} & 1.0  & 1.0 & 1.0 \\
            \hline
            \textbf{Type II} & 1.0 & 1.0 & 1.0 \\
            \hline
            \textbf{Type III} & 0.1703 & 0.3247 & 0.6890 \\
            \hline
            \textbf{Type IV} & 0.0291 & 0.1032 & 0.5443\\
            \hline
            \textbf{Mean} & 0.5498 & 0.6069 & 0.8083\\
            \bottomrule       
        \end{tabular}
    }
    \label{tab:rule_based_vs_poj_evl}
\end{table}

\revision{As shown in Table~\ref{tab:rule_based_vs_poj_evl}, our method consistently outperforms both the Vanilla embeddings and the rule-based augmentation across all categories. Notably, for the challenging Type III and Type IV, POJ-Evl achieves substantial improvements (0.6890 and 0.5443, respectively) compared to the modest gains from rule-based augmentation, demonstrating the effectiveness of our systematic design for functional diversity.}

\section{Conclusion}
Existing datasets for code intelligence tasks largely focus on syntactic similarity when evaluating code pairs, with an overrepresentation of syntactically similar (Type II) or dissimilar (Type III) examples. This imbalance limits their ability to assess functional consistency in more complex scenarios, such as functionally equivalent code with significant syntactic differences (Type II) or functionally inconsistent code with superficial syntactic resemblance (Type IV).
To address this problem, we proposed  \lzhmark{functionality-oriented code self-evolution}, which generates diverse code variations that balance syntactic and semantic differences while preserving functional consistency where applicable. By creating a more comprehensive dataset, our framework enables embedding models to capture deeper semantic understanding.
%
To validate the advancements of our proposed framework, we evaluated our evolved datasets using embedding models on three downstream tasks: code clone detection, code functional consistency, and code retrieval. Experimental results show a significant improvement in the model's ability to represent code semantics, even across different code datasets, demonstrating the effectiveness and superiority of our code self-evolution.
\revision{In future work, we plan to explore extending our framework to support multi-language scenarios and incorporate dynamic analysis techniques. These directions aim to further enhance the robustness and accuracy of code understanding and evaluation in increasingly complex and diverse programming environments.}

\section*{\revision{Limitations}}
\revision{Our functionality-oriented code self-evolution framework achieves consistent gains across downstream tasks, yet certain limitations remain. First, although it has been validated on widely used benchmarks (e.g., CodeXGlue, HumanEval), its generalizability could be further improved by expanding evaluation to more languages and programming paradigms. Second, variations in LLM instruction design may affect the results, indicating potential benefits from further optimization. Future work will address these aspects to strengthen the framework's applicability.}

\section*{Data Availability}
To support the reproducibility, extensibility, and practical utility of our proposed framework, we have made the evolved code datasets and implementation code available. These resources can be accessed at: \href{https://drive.google.com/drive/folders/1qO9n8az4AgNfL8Cg_0nXQeYOn19dfltg?usp=drive_link}{\textit{Functionality-Oriented Code Self-Evolution}}. We believe this contributes a valuable benchmark for future research on functional code understanding.

\section*{Acknowledgements}
This work was supported by the National Natural Science Foundation of China (62306344, 62276279) and Guangdong Basic and Applied Basic Research Foundation (2024A1515010253, 2024B1515020032).






\bibliographystyle{elsarticle-num} 
\bibliography{ref}






\newpage
\appendix  
\section{}
In this section, we provide additional experimental results and the code evolution instructions for our proposed code self-evolution method and the task alignment instructions used in downstream task experiments.
Additionally, we provide more details about our evolved datasets.
\subsection{Additional Experiments} \label{appendix-bert}
As shown in Table~\ref{tab:test_evolved_dataset2}, we present fine-tuning results for BERT-level models (CodeBert~\cite{codebert} and Unxicoder~\cite{guo-etal-2022-unixcoder}) based on the POJ-Evl dataset. The experimental outcomes show that both models are consistent with M3~\cite{BGE-M3}, demonstrating that POJ-Evl offers a significant increase in overall complexity compared to the POJ dataset, thereby highlighting the superiority of our proposed code evolution method.
\begin{table}[ht]
    \caption{Dataset Effectiveness Comparison on the Bert-based models in Terms of F1 Score.} 
    \centering
        \resizebox{0.7\textwidth}{!}
        {
        \begin{tabular}{ l | c | c | c }
            \toprule
             & \textbf{CodeBert} & \textbf{CodeBert\_poj} & \gt{\textbf{CodeBert\_evl}} \\ 
             \hline
            \textbf{Mean} & 0.5010 \rt{(+0.130)} & 0.5 \rt{(+0.131)} & \gt{0.631} \\
            \midrule       
            & \textbf{Unixcoder} & \textbf{Unixcoder\_poj} & \gt{\textbf{Unixcoder\_evl}} \\ 
             \hline
            \textbf{Mean} & 0.5 \rt{(+0.1229)} & 0.4146 \rt{(+0.2083)} & \gt{0.6229} \\
            \bottomrule
        \end{tabular}
    }
    \label{tab:test_evolved_dataset2}
\end{table}

\subsection{Details of the Instructions}
\subsubsection{Instructions for Functionality-Oriented Code Self-Evolution}
\label{sec:instru1}
As shown in Table~\ref{tab:evol_instructions} and Table~\ref{tab:evol_typeIII}, we provide a detailed list of four code self-evolution instructions.

\begin{table}[H]
    \caption{
     Instructions for code self-evolution.
    } 
    \centering
    \tabcolsep=0.27cm
    \renewcommand{\arraystretch}{1} 
    \begin{tabularx}{\textwidth}{c|X}
        \hline
        \textbf{Types}  & \textbf{Instructions}  \\ 
        \hline
        Type I & 
        I want you act as a Code Rewriter.\textbackslash r\textbackslash n 
        Your objective is to rewrite a given piece of code through a series of random code perturbations while maintaining its functionality.\textbackslash r\textbackslash n 
        Principle: The perturbations should be RANDOM and DIVERSE. You can select ONE or MULTIPLE perturbation methods for combination. 
        You can refer to, but not be limited to, the following code perturbations: Change Function/Variable Names, Change Docstring, and Dead Code.\textbackslash r\textbackslash n 
        For usability, you SHOULD provide the correct, runnable code without additional analysis.\textbackslash r\textbackslash n \\
        \hline
        Type II & 
        I want you act as a Code Rewriter.\textbackslash r\textbackslash n \
        Your objective is to rewrite a given piece of code to ensure the functionality remains the same, but the syntax, structure, or implementation details differ significantly.\textbackslash r\textbackslash n \
        Principle: The rewritten code should be semantically equivalent to the original, but it should not resemble the original in terms of coding style, variable names, or structure.\textbackslash r\textbackslash n \
        For usability, you SHOULD directly provide the runnable rewritten code while ensuring it passes all possible functional tests of the original code.\textbackslash r\textbackslash n 
        \\ \hline
        Type III & I want you act as a Code Rewriter.\textbackslash r\textbackslash n\ Your objective is to 1. generate a more complex code version which performs absolutely different functions/intentions from given code and 2. generate a set of test inputs.\textbackslash r\textbackslash n \
        You SHOULD make sure the given code and new code have different functionalities/intentions. You can generate new code based on the following Code Evolution Instructions, but do not limit yourself to the instructions:\textbackslash r \textbackslash n
        \\ \hline
        Type IV & I want you act as a Code Rewriter.\textbackslash r\textbackslash n\
Your objective is to rewrite a given piece of code in a way that the resulting code appears similar in structure but performs absolutely different core functionality.\textbackslash r\textbackslash n \
The rewritten code should implement a fundamentally different function from the original code, rather than simply differing in naming conventions, printing operations, or other aspects that do not affect the core functionality.\textbackslash r\textbackslash n \
You SHOULD directly provide an executable code.\textbackslash r\textbackslash n Keep the format of the output codes consistent with the original codes.\textbackslash r\textbackslash n
        \\ \bottomrule
    \end{tabularx}

    \label{tab:evol_instructions}
\end{table}

\begin{table*}[!t]
    \caption{
     Instructions for Type III.
    } 
    \centering
    \renewcommand{\arraystretch}{1} 
    \begin{tabularx}{\textwidth}{ c | X } 
        \hline
         & \textbf{Evolving Instructions}  \\ 
        \hline
        1 & Choose your preferred way to change the given code's intent and generate new code.\textbackslash r\textbackslash n \\
        \hline
        2 & Please add one more constraint/requirements.\textbackslash r\textbackslash n \\
        \hline
        3 & Please replace general concepts with more specific concepts.\textbackslash r\textbackslash n \\
        \hline
        4 & You can generate entirely different new code by explicitly requesting multi-step reasoning involving different functionalities/intents from the given code.\textbackslash r\textbackslash n \\
        \hline
        5 & You can generate entirely different new codes by increasing the depth and breadth of the inquiry involving different functionalities/intents from the given code.\textbackslash r\textbackslash n \\
        \hline
    \end{tabularx}

    \label{tab:evol_typeIII}
\end{table*}

\subsubsection{Instructions for Task Alignment Self-Evolution} \label{sec:instru2}
Additionally, we employed task alignment (ICL and Instruction) for all code tasks. The task instructions used are shown in Table~\ref{tab:instructions}.
\begin{table*}[ht] \large
    \caption{
    Task instructions for task alignment.
    } 
    \centering
    \renewcommand{\arraystretch}{1.35}
    \resizebox{\textwidth}{!}
    {
        \begin{tabular}{ l |  c }
            \toprule
            \textbf{Task}  & \textbf{Instruction}  \\ 
            \hline
            Code Clone & Given a code query, detect similar code clones.\\
            Functional Consistency & Given a code query, identify other code snippets with the same functional consistency.\\
            Code Retrieval & Given a code query, retrieve relevant code snippets that address the same problem as the query. \\
            \bottomrule            
        \end{tabular}
    }
    \label{tab:instructions}
\end{table*}

\subsection{More Details about Evolved Datasets}
In order to demonstrate the effectiveness and advancement of our proposed data synthesis framework, we have developed evolved versions based on the code dataset, which we refer to as the POJ-Evl and HumanEval-Evl datasets.
From a single code instance, we can generate four code variants that have evolved both syntactically and semantically.
\textit{It is important to note that the proposed code evolution methods can be applied to any code dataset, not limited to the POJ-104 and HumanEval datasets.}

\subsection{\revision{Power Analysis for Downstream Tasks} \label{sec:power_ananlysis}}
\revision{To further support the statistical reliability of our reported improvements, we conducted power analyses for the +Task Alignment $\rightarrow$ +POJ-Evl comparisons across all downstream tasks. 
For each task, results are based on five independent runs with different random seeds, using paired $t$-tests to compute effect sizes (Cohen's $d$), statistical power, and bootstrap 95\% confidence intervals (CIs) for the mean differences. 
The outcomes, summarized in Table~\ref{tab:power_analysis}, demonstrate consistently large effect sizes and high power for the majority of tasks, indicating that the observed improvements are unlikely to be due to random variation.}

\begin{table}[H]
\centering
\caption{\revision{Summary of power analysis results for +Task Alignment $\rightarrow$ +POJ-Evl comparisons across downstream tasks. Reported as paired Cohen’s $d$, statistical power, and bootstrap 95\% confidence interval (CI) for the mean difference, based on five independent runs with different random seeds.}}
\resizebox{\textwidth}{!}{
\begin{tabular}{lccc}
\toprule
\textbf{Task} & \textbf{Paired Cohen's $d$} & \textbf{Statistical Power} & \textbf{Bootstrap 95\% CI (Mean Diff.)} \\
\midrule
Code Retrieval & 0.9015 & 0.3402 & [0.1900, 2.0140] \\
Code Functional Consistency & 23.6665 & 1.0000 & [0.5028, 0.5368] \\
Code Clone Detection & 1.9694 & 0.9009 & [0.0081, 0.0184] \\
\bottomrule
\end{tabular}
}
\label{tab:power_analysis}
\end{table}

\end{document}